\documentclass[10pt,aps,prd,noshowpacs,eqsecnum,nofootinbib,twocolumn]
{revtex4-1}
\usepackage{graphicx}
\usepackage{amsmath,amsthm,amssymb}
\usepackage{enumerate}

\begin{document}
\title{Revised 
Criteria for Stability in the General Two-Higgs Doublet Model}

\author{Yithsbey Giraldo and Larry Burbano}

\affiliation{Departamento de F\'\i sica, Universidad de Nari\~no, A.A. 1175, 
San 
Juan de Pasto, 
Colombia}

 \email[E-mail:]{yithsbey@gmail.com}

\begin{abstract}
We will revise one of the methods given in the literature to determine the 
necessary and sufficient conditions that the parameters must satisfy to have a 
stable scalar potential in the general two-Higgs doublet model. We will give a 
procedure that facilitates finding the conditions for stability of a scalar 
potential. The stability guarantees that the scalar potential has a global 
minimum,  that is, the potential is bounded from below, which is a necessary 
condition to implement the spontaneous gauge-symmetry breaking in the models.
\end{abstract}

\maketitle

\section{Revised criteria for stability}

We obtain the stationary points of $J_4(\mathbf k)$ using Eq.~(46) of 
Ref.~\cite{Maniatis:2006fs}  as 
follows:
\begin{equation}
\label{1}
 (E-u)\mathbf k=-\boldsymbol \eta,\qquad\textrm{with}\quad |\mathbf k|=1,
\end{equation}
where $|\mathbf k|<1$ for the case $u=0$.
Now, suppose we find two solutions $\mathbf p$ and $\mathbf q$ with their 
respective 
Lagrange multipliers $u_p$ and $u_q$ such that
\begin{equation}
\label{2}
 \begin{split}
   (E-u_p)\mathbf p&=-{\boldsymbol \eta},\\
    (E-u_q)\mathbf q&=-\boldsymbol\eta,
 \end{split}
\end{equation}
where
 \begin{equation}
 \label{3}
  |\mathbf p|=1, \quad\quad  |\mathbf q|=1 \quad\textrm{and}\quad u_p\neq u_q.
 \end{equation}
Let us evaluate the function~$J_4(\mathbf k)$ at these stationary points
 
 \begin{equation}
 \label{4}
 \begin{split}
 J_4(\mathbf p)&=u_p+\eta_{00}+\boldsymbol\eta^T\mathbf p,\\
 J_4(\mathbf q)&=u_q+\eta_{00}+\boldsymbol\eta^T\mathbf q.
 \end{split}
 \end{equation}
Given that the matrix $E$ is symmetric, from Eq.~\eqref{2},  we have 
\begin{equation}
\label{5}
 (u_q-u_p)\mathbf p^T\mathbf q=\boldsymbol\eta^T(\mathbf p-\mathbf q),
\end{equation}
and taking into account~\eqref{4}, we obtain
\begin{equation}
\label{6}
 J_4(\mathbf p)-J_4(\mathbf q)=(u_p-u_q)(1-\mathbf p^T\mathbf q).
\end{equation}
The product $\mathbf p^T\mathbf q=|\mathbf p||\mathbf 
q|\cos\theta=\cos\theta<1$. According to~\eqref{3}, the case $\theta=0$  
implies 
that $\mathbf p=\mathbf q $, and from~\eqref{5}, we deduce $u_p=u_q$, which 
contradicts 
the 
assumed in~\eqref{3}. Further, the inequality $\mathbf p^T\mathbf q <1$ is 
immediately 
satisfied if $|\mathbf p|<1$. Therefore, in any case, it is true that the 
factor 
$(1-\mathbf p^T\mathbf q)>0$, 
and consequently, from~\eqref{6}, we conclude that
\begin{equation}
\label{7}
u_p<u_q\Longleftrightarrow J_4(\mathbf p)<J_4(\mathbf q).
\end{equation}
The result~\eqref{7} is quite useful because it makes it easier to find the 
conditions 
of the parameters to have a stable scalar potential. The process would be as 
follows: 
compute all the ``regular'' Lagrange multipliers $\{u_i\} $~$ (i\leq 6)$, by 
solving 
equation~(52) of Ref.~\cite{Maniatis:2006fs}. Include in this set the 
``exceptional'' solutions 
$\{\mu_j\}$~$(j\leq 3)$, by solving the equation $\det(E-u)=0$, omitting the 
values 
$\mu_j$ for which the corresponding $\eta_j\neq 0$, on the basis that E is 
diagonal~(as 
you can see from~\eqref{1}). 
Finally, consider $u=0$ for 
solutions within the sphere $|\mathbf k| <1$, and with them, form the set
$S=\{u_i, 
\mu_j,0\}$, which has at most ten elements.

The result~\eqref{7} suggests taking the smallest value of $S$ to establish a 
stable scalar potential. Since the values of $S$ are in general free parameters, 
let us 
assume that 
each one of 
them is the lowest value.

If the smallest value is a regular solution $\{u_i\} $, immediately impose the 
condition $J_4(\mathbf p_i)>0$, that is, $f(u_i)>0$ (Eq.~(51) in 
Ref.~\cite{Maniatis:2006fs}), that 
which, according to the result~\eqref{7}, would guarantee the stability of the 
scalar 
potential. Conditions coming from regular solutions are necessary. Let us keep 
the 
regular 
solutions in $S$.

If the smallest value is an exceptional solution, $\{\mu_j\} $, you must first 
verify 
that it gives a valid stationary point, that is, $f'(\mu_j)\geq0$. If this is 
not right, 
you can discard this value from the set $S$. In the case of being satisfied, 
impose the
condition $f(\mu_j)>0$, which would guarantee the stability of the potential 
according to the result~\eqref{7}. The conditions arising from the exceptional 
solutions may not be necessary since the inequality $f'(\mu_j)\geq0$ is not 
always 
satisfied. Similarly, if the smallest value of $S$ is 0, you should check first 
that 
$f'(0)>0$; if not, discard this value from $S$. If it is satisfied, set the 
condition $f(0)>0$ to ensure the stability of the scalar potential.

Values of $S$ that, given their structure, cannot be the smallest, are discarded 
if 
the 
lowest value gave a valid stationary point~(according to the result~\eqref{7}). 
Otherwise, they should be analyzed. 

So far, the conditions above give stability in a ``strong'' sense. If 
for one of the cases above we have $f(u)=0$, proceed as indicated in 
Ref.~\cite{Maniatis:2006fs,Nagel:2004sw},  
considering,  in this case,  $J_2(\mathbf k)$,  
which would guarantee the stability of the scalar potential in the weak or 
marginal 
sense. For 
the remaining stationary points, it follows that $J_4(\mathbf k)>0$,  as stated 
in~\eqref{7}.

Finally, we build the set
\begin{equation}
 I=\{\textrm{values not discarded from $S$}\},
\end{equation}
from which we obtain the sufficient conditions to guarantee the stability of the 
scalar 
potential. Let us apply the results above to a particular model.

\section{Example: Stability for THDM}
\label{thdm}
 Let us analyze the 
two-Higgs-doublet model~(THDM) of Gunion et al., with the Higgs 
potential given in Eq.(79) of Ref.~\cite{Maniatis:2006fs}. After  examining the 
potential, the 
corresponding Lagrange multipliers, including 0, which could result in possible 
stability 
conditions, give the following set:
{\small
\begin{equation}
\label{9}
\begin{split}
S=&\left\{u_1=\frac{1}{4}(2\lambda_1-\lambda_4), 
u_2=\frac{1}{4}(2\lambda_2-\lambda_4),\right.
u_3=0,\mu_4=\frac{1}{4}\left(\kappa-\lambda_4\right),
\\
&\left.\mu_5=\frac{1}{8}
\left(-2\lambda_4+\lambda_5+\lambda_6+\sqrt{(\lambda_5-\lambda_6)^2+\lambda_7^2}
\right)\right\},
\end{split}
\end{equation}}
where $\kappa=\frac{1}{2}\left(\lambda_5+\lambda_6-\sqrt{
(\lambda_5-\lambda_6)^2+\lambda_7^ 
2} 
\right)$. The first two parameters are the regular Lagrange multipliers, and 
the 
last two are the appropriate exceptional solutions in $S$. Note that 
$\mu_4<\mu_5$, 
but we still cannot discard $\mu_5$ since we must first check if 
$f'(\mu_4)\geq0$. The 
global minimum of 
$J_4(\mathbf k)$ occurs where the minimum valid value of $S$ is.
\begin{enumerate}[(i)]
\item If $u_1$ is the smallest value of $S$ in~\eqref{9}, then
\begin{equation}
\label{10}
 f(u_1)>0\Longrightarrow \lambda_1+\lambda_3>0.
\end{equation}

\item If $u_2$ is the smallest value of $S$ in~\eqref{9}, then
\begin{equation}
\label{11}
 f(u_2)>0\Longrightarrow \lambda_2+\lambda_3>0.
\end{equation}
Since $u_1$ and $u_2$ are regular solutions, the inequalities~\eqref{10} 
and~\eqref{11} 
are 
necessary.
\item 
%
\begin{equation}
\label{12}
\textrm{If}\quad u_3=0<u_1,u_2,\mu_4,\mu_5,
\end{equation}
we can observe that
\begin{equation}
 f'(u_3)=\frac{4u_1u_2}{(u_1+u_2)^2}>0,
\end{equation}
so $u_3$ is not discarded. Taking into account the 
inequalities~\eqref{10} and~\eqref{11} in $f(u_3)$, we have
\begin{widetext}
 \begin{equation} 
f(u_3)=\frac{\left[-\lambda_4-2\lambda_3+2\sqrt{
(\lambda_1+\lambda_3)(\lambda_2+\lambda_3)}\right]
\left[\lambda_4+2\lambda_3+2\sqrt{
(\lambda_1+\lambda_3)(\lambda_2+\lambda_3)}\right]}{8(u_1+u_2)}>0,
\end{equation}
\end{widetext}
and from Eq.~\eqref{12} we can show that the factors $u_1+u_2>0 $ and 
$-\lambda_4-2 
\lambda_3+2\sqrt{(\lambda_1+\lambda_3) (\lambda_2+\lambda_3)}>0$; therefore
\begin{equation}
 \lambda_4>-2\lambda_3-2\sqrt{
(\lambda_1+\lambda_3)(\lambda_2+\lambda_3)}.
\end{equation}

\item  
%
\begin{equation}
\label{16}
 \textrm{If}\quad \mu_4<u_1,u_2,u_3,\mu_5,
\end{equation}
then
\begin{equation}
\label{17}
f'(\mu_4)=\frac{(2\lambda_1-\kappa)(2\lambda_2-\kappa)}{
(\lambda_1+\lambda_2-\kappa)^2}>0
\end{equation}
because of inequalities $(2\lambda_1-\kappa)>0$, $(2\lambda_2-\kappa)>0$ and 
$(\lambda_1 
+\lambda_2-\kappa)>0$ derived from the Eq.~\eqref{16}. So, the Lagrange 
multiplier 
$\mu_4$ 
must be included in the set $I$. Besides,
\begin{widetext}
 \begin{equation} 
f(\mu_4)=\frac{\left[-\kappa-2\lambda_3+2\sqrt{
(\lambda_1+\lambda_3)(\lambda_2+\lambda_3)}\right]
\left[\kappa+2\lambda_3+2\sqrt{
(\lambda_1+\lambda_3)(\lambda_2+\lambda_3)}\right]}{
4(\lambda_1+\lambda_2-\kappa)}>0,
\end{equation}
\end{widetext}
and using~\eqref{16}, we can show that the factors 
$(\lambda_1+\lambda_2-\kappa)>0$ and {\small
$-\kappa-2\lambda_3+2\sqrt{(\lambda_1+\lambda_3)(\lambda_2+\lambda_3)}>0$}; 
therefore
\begin{equation}
 \kappa>-2\lambda_3-2\sqrt{
(\lambda_1+\lambda_3)(\lambda_2+\lambda_3)}.
\end{equation}
\end{enumerate}
So the Lagrange multiplier $\mu_5$ is not considered since $\mu_5>\mu_4$. 

In short, for the THDM to be stable, the following conditions on the parameters 
are {\it sufficient}
\begin{widetext}
 \begin{equation}
  \lambda_1+\lambda_3>0,\quad \lambda_2+\lambda_3>0,\quad
\lambda_4,\kappa>-2\lambda_3-2\sqrt{(\lambda_1+\lambda_3)(\lambda_2+\lambda_3)}.
\end{equation}
\end{widetext}

\section{Conclusions}
We can see that the application of the result~\eqref{7} is essential to get a 
consistent 
model and be able to derive sufficient conditions to have a stable scalar 
potential. It 
allows us to identify either necessary conditions~(for regular solutions) 
or
conditions that may not be  necessary, coming from exceptional 
solutions~(including 0). 
Both conditions generate sufficient inequalities that guarantee the stability of 
a 
scalar 
potential. As an example, we can appreciate it, in the expression (152) of 
Ref.~\cite{Maniatis:2006fs}, where $u_2<u_1,u_3$, so for stability conditions, 
only $u_2$ is 
considered. 
In this sense, it may happen that some Lagrange multipliers, although not 
being the smallest values, must be taken into account for stability conditions. 
 You can appreciate it from Gunion's potential in Sect.~\ref{thdm}~(Eq.~(79) 
of 
Ref.~\cite{Maniatis:2006fs}), since if
$\mu_4$ were not a valid stationary point, we would have 
had to analyze $\mu_5$. In that way, we can reduce the number of sufficient 
conditions arising from exceptional solutions~(including
0) provided that $f'(\mu_j)<0$~(or~$f'(0)\leq0$).

\end{document}